# Sharing Neurophysiology Data from the Allen Brain Observatory: Lessons Learned

Saskia E. J. de Vries[†]*, Joshua H. Siegle[†]*, Christof Koch[†]

[†]Allen Institute, Seattle, WA 98109

*Equal contributors and corresponding authors: saskiad@alleninstitute.org, joshs@alleninstitute.org

**Abstract**

Making all data for any observation or experiment openly available is a defining feature of empirical science (e.g., *nullius in verba*, the motto of the Royal Society). It enhances transparency, reproducibility, and societal trust. While embraced in spirit by many, in practice open data sharing remains the exception in contemporary systems neuroscience. Here, we take stock of the Allen Brain Observatory, an effort to share data and metadata associated with surveys of neuronal activity in the visual system of laboratory mice. The data from these surveys have been used to produce new discoveries, to validate computational algorithms, and as a benchmark for comparison with other data, resulting in over 100 publications and preprints to date. We distill some of the lessons learned about open surveys and data reuse, including remaining barriers to data sharing and what might be done to address these.

## Introduction

Why share data? The central nervous system is among the most complex organs under investigation. Accordingly, the tools to study it have become intricate and costly, generating ever-growing torrents of data that need to be ingested, quality-controlled, and curated for subsequent analysis. Not every lab has the financial or personnel resources to accomplish this. Moreover, while many scientists relish running experiments, others find their passion in analysis. Data collection requires a different skillset than analysis, especially as the field demands more comprehensive and higher-dimensional datasets which, in turn, necessitate more advanced analytical methods and software infrastructure to interpret. A scientific ecosystem in which data is extensively shared and reused would give researchers more freedom to focus on their favorite parts of the discovery process.

Sharing data brings other benefits as well. It increases the number of eyes on each dataset, making it easier to spot potential outlier effects (Button et al., 2013). It encourages meta-analyses that integrate data from multiple studies, providing the opportunity to reconcile apparently contradicting results or expose the biases inherent in specific analysis pipelines (Botvinik-Nezer et al., 2020; Mesa et al., 2021). It also gives researchers a chance to test hypotheses on existing data, refining and updating their ideas before embarking on the more costly process of running new experiments.

Without a doubt, re-analysis of neurophysiology data has already facilitated numerous advances. Electrophysiological recordings from non-human primates, which require tremendous dedication to collect, are often reused in multiple high-impact publications (Churchland et al., 2010; Murray et al., 2014). In the domain of hippocampal physiology, the lab of György Buzsáki maintains a databank of recordings from more than 1000 sessions from freely moving rodents (Petersen et al., 2020). And data from "calibration" experiments, in which activity of individual neurons is monitored via two modalities at once, have been extremely valuable for improving data processing algorithms (GENIE Project, 2015; Henze et al., 2009; Huang et al., 2021; Neto et al., 2016). A number of these datasets have been shared via the website of CRCNS (Teeters et al., 2008), a far-sighted (but now defunct) organization focused on aggregating data from multiple species within the same searchable database. More recently, an increasing number of researchers are choosing to make data public via generalist repositories such as Figshare, Dryad, and Zenodo. As data can be hosted on these repositories for free, they greatly lower the barriers to sharing. However, because there are no restrictions on the data format or level of documentation, learning how to analyze open datasets can take substantial effort, and scientists are limited in their ability to perform meta-analyses across datasets.

Since the time of its founding, the Allen Institute has made open data one of its core principles. Specifically, it has become known for generating and sharing *survey* datasets within the field of neuroscience, taking inspiration from domains such as astronomy where such surveys are common (York et al., 2000). The original Allen Mouse Brain Atlas (Lein et al., 2007) and subsequent surveys of gene expression (Bakken et al., 2016; Hawrylycz et al., 2012; Miller et al., 2014), mesoscale connectivity (Harris et al., 2019; Oh et al., 2014), and in vitro firing patterns (Gouwens et al., 2019) have become essential resources across the field. These survey datasets are (1) collected in a highly standardized manner with stringent quality controls, (2) create a volume of data that is larger than typical individual studies within their particular disciplines, and (3) are collected without a specific hypothesis to facilitate a diverse range of use cases.

Starting a decade ago, we began planning the first surveys of *in vivo* physiology with single-cell resolution (Koch and Reid, 2012). Whereas gene expression and connectivity are expected to change relatively slowly, neural responses in awake subjects can vary dramatically from moment to moment, even during apparently quiescent periods (McCormick et al., 2020). Therefore, an *in vivo* survey of neural activity poses new challenges, requiring many trials and sessions to account for both intra- as well as inter-subject variability. We first used 2-photon calcium imaging and later Neuropixels electrophysiology to record spontaneous activity and evoked responses in visual cortex and thalamus of awake mice that were passively exposed to a wide range of visual stimuli (known as “Visual Coding” experiments). The inclusion of a large number of subjects, highly standardized procedures, and rigorous quality control criteria distinguished these surveys from typical small-scale neurophysiology studies. More recently, the Institute carried out surveys of single-cell activity in mice performing a visually guided behavioral task (known as “Visual Behavior” experiments). In all cases, the data was shared even before we published our own analyses of them. We reflect here on the lessons learned on the challenges of data sharing and reuse in the neurophysiology space. Our primary takeaway is that the widespread mining of our publicly available resources demonstrates a clear demand for open neurophysiology data and points to a future in which data reuse becomes more commonplace. However, more work is needed to make data sharing and reuse practical (and ideally the default) for all laboratories practicing systems neuroscience.

## Overview of the Allen Brain Observatory

The Allen Brain Observatory consists of a set of standardized instruments and protocols designed to carry out surveys of cellular-scale neurophysiology in awake brains (de Vries et al., 2020; Siegle et al., 2021a). Our initial focus was on neuronal activity in the mouse visual cortex (Koch and Reid, 2012). Vision is the most widely studied sensory modality in mammals, but much of the foundational work is based on recordings with hand-tuned stimuli optimized for individual neurons, typically investigating a single area at a time (Hubel and Wiesel, 1998). The field has lacked the sort of unbiased, large-scale surveys required to rigorously test theoretical models of visual function (Olshausen and Field, 2005). The laboratory mouse is an advantageous model animal given the

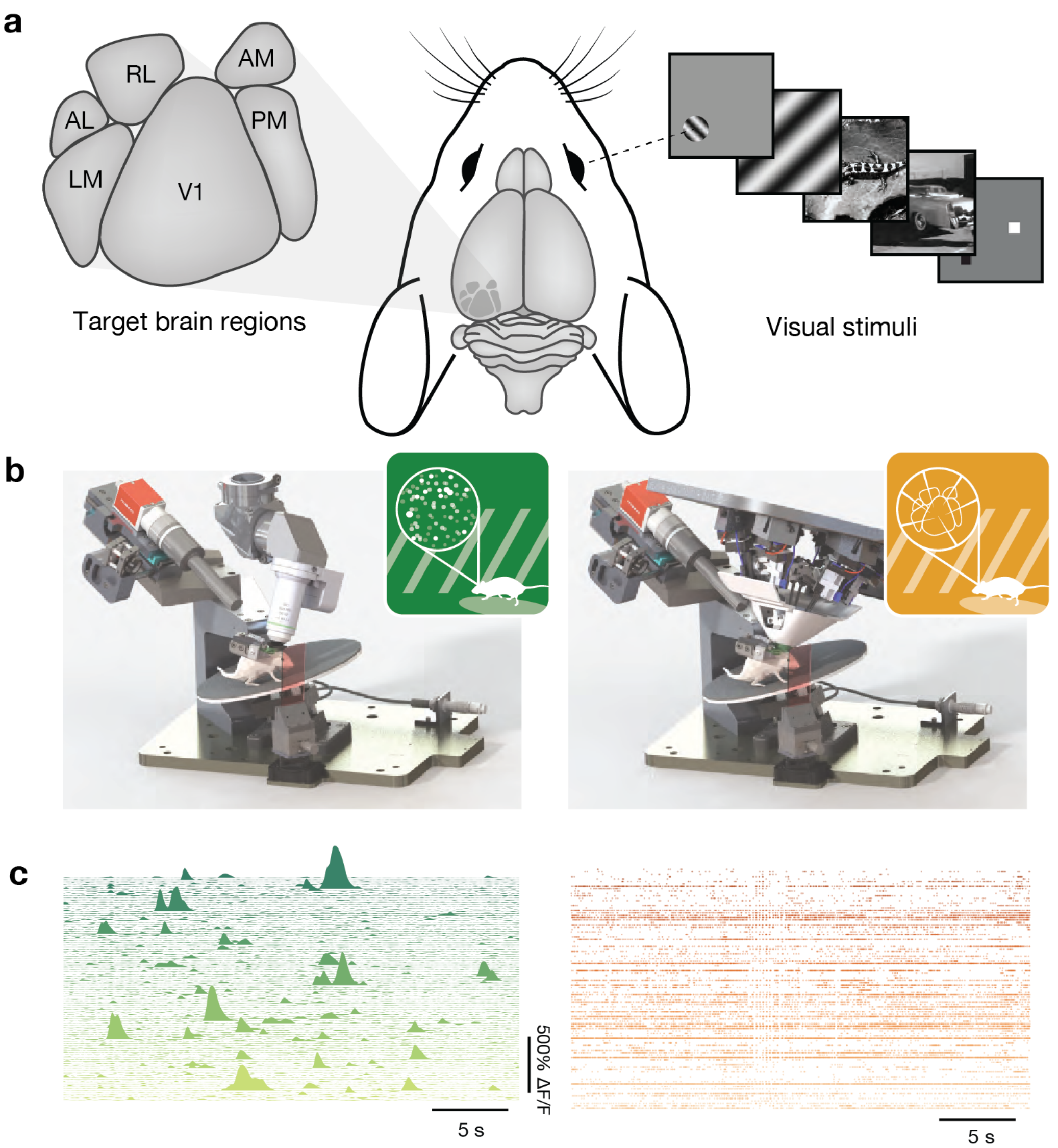

**d**

| | 2P | Electrophysiology | Combined dataset size |
|---|---|---|---|
| **Raw** | Movies | Voltage traces | 514 experiments |
| **Processing** | Motion correction | Median subtraction | 1700 probes or planes |
| **Annotation** | Segmentation, event detection | Spike sorting | 103,261 cells or units |
| **Extraction** | Peri-stimulus ΔF/F traces | Peri-stimulus spike trains | 11.4M presentations |
| **Quantification** | Response magnitude | Firing rate | 5M metrics |

**Figure 1 – Overview of Allen Brain Observatory Visual Coding datasets.** (a) Target brain regions and example visual stimuli. (b) Standardized rigs for 2-photon calcium imaging (left) and Neuropixels electrophysiology (right). (c) Example ΔF/F traces or spike rasters for 100 simultaneously recorded neurons from each modality. Both are extracted around one presentation of a 30 s natural movie stimulus. (d) Dataset size after different stages of analysis.

extensive ongoing work on mouse cell types (BRAIN Initiative Cell Census Network et al., 2021; Tasic et al., 2018; Yao et al., 2021; Zeisel et al., 2015), as well as access to a well-established suite of genetic tools for observing and manipulating neural activity via driver and reporter lines or viruses (Gerfen et al., 2013; Madisen et al., 2015).

Our 2-photon calcium imaging dataset (Allen Institute MindScope Program, 2016) leveraged transgenic lines to drive the expression of GCaMP6 (Chen et al., 2013) in specific populations of excitatory neurons (often constrained to a specific cortical layer) or GABAergic interneurons. In total, we recorded activity from over 63,000 neurons across 6 cortical areas, 4 cortical layers, and 14 transgenic lines (**Figure 1**). The Neuropixels electrophysiology dataset (Allen Institute MindScope Program, 2019) used silicon probes (Jun et al., 2017) to record simultaneously from the same 6 cortical areas targeted in the 2-photon dataset, as well as additional subcortical regions (Durand et al., 2022). While cell type–specificity was largely lost, transgenic lines did enable optotagging of specific inhibitory interneurons. The Neuropixels dataset included recordings from over 40,000 units passing quality control across more than 14 brain regions and 4 mouse lines (**Figure 1**). In both of these surveys, mice were passively exposed to a diverse and overlapping array of visual stimuli. Subsequent surveys of neural activity in mice performing a behavioral task are not discussed here, as it is too soon to begin evaluating their impact on the field.

## Approach to data distribution

Once the data was collected, we wanted to minimize the friction required for external groups to access it and mine it for insights. This is challenging! Providing unfettered access to the data can be accomplished by providing a simple download link; yet, unless the user understands what is contained in the file and has installed the appropriate libraries for parsing the data, its usefulness is limited. At the other extreme, a web-based analysis interface that does not require any downloading or installation can facilitate easy data exploration, but this approach has high up-front development costs and imposes limitations on the analyses that can be carried out.

These conflicting demands are apparent in our custom tool, the AllenSDK, a Python package that serves as the primary interface for downloading data from these surveys, as well as other Allen Institute resources. In the case of the Allen Brain Observatory, the AllenSDK provides wrapper functions for interacting with the Neurodata Without Borders (NWB) files (Rübel et al., 2022; Teeters et al., 2015) in which the data is stored. Intuitive functions enable users to search metadata for specific experimental sessions and to extract the relevant data assets. Whereas our 2-photon calcium imaging survey was accompanied by a dedicated web interface that displayed summary plots for every cell and experiment (observatory.brain-map.org/visualcoding), we discontinued this practice because most users preferred to directly access the data in their own analysis environment.

One challenge with sharing cellular neurophysiology data is that it includes multiple high-dimensional data streams. Many other data modalities (e.g., gene expression) can be reduced to a derived metric and easily shared in a tabular format (e.g., *cell-by-gene* table). In contrast, neurophysiological data is highly varied, with analysts taking different approaches to both data processing (e.g., spike sorting or cell segmentation) and analysis. We aimed to share our data in a flexible way to facilitate diverse use cases. Accordingly, we provided either spike times or fluorescence traces, temporally aligned stimulus information, the mouse's running speed and pupil tracking data, as well as intermediate, derived data constructs, such as ROI masks, neuropil traces, and pre- and post- de-mixing traces for two-photon microscopy, and waveforms across channels for Neuropixels. All are contained within the NWB files. In addition, we provided access to the more cumbersome, terabyte-scale raw imaging movies and voltage traces via an Amazon S3 bucket for users focused on data processing algorithms (**Figure 2**).

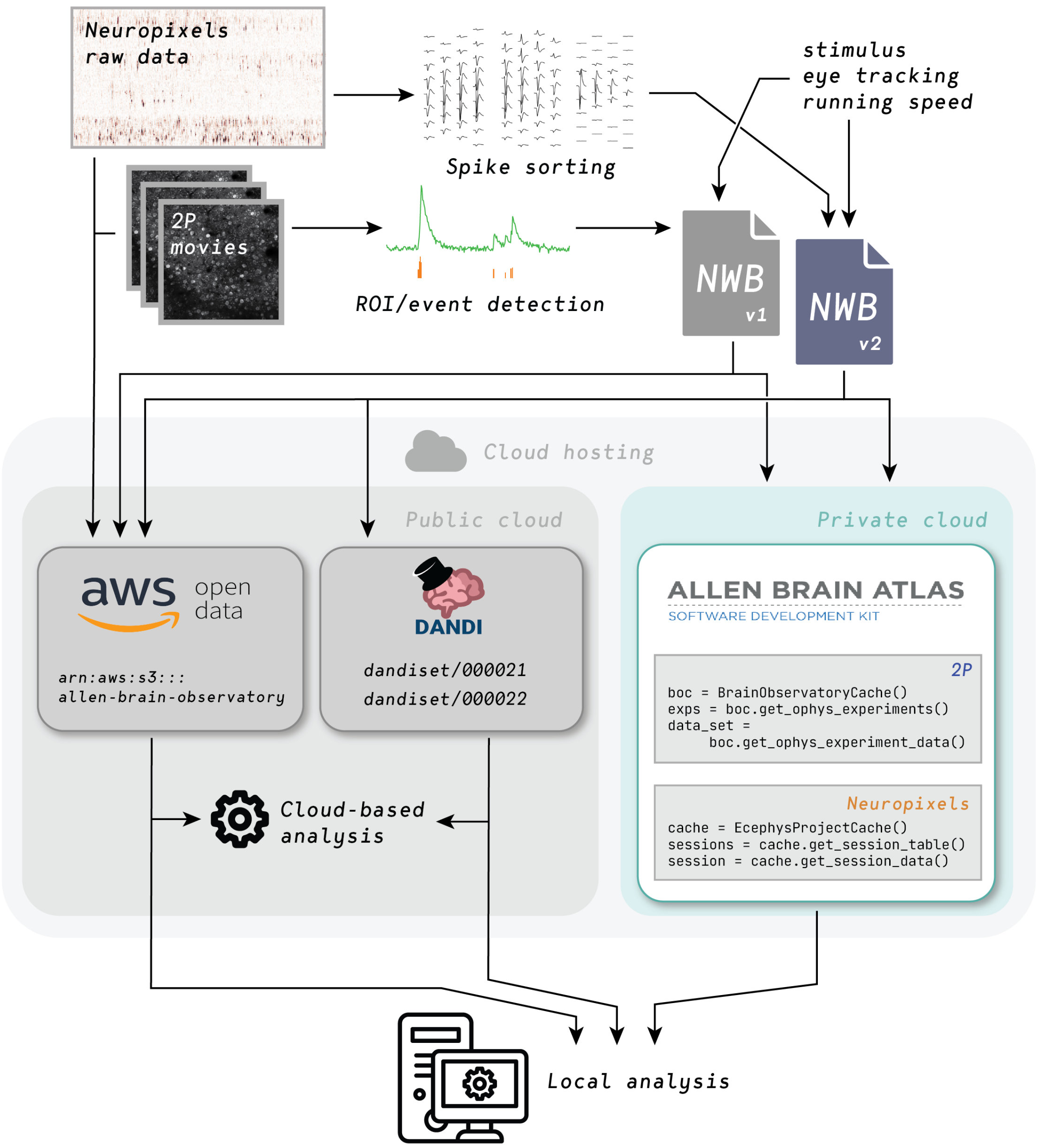

**Figure 2 – Distributing data from Allen Brain Observatory Visual Coding experiments.** Raw data is acquired and processed at the Allen Institute, combined with metadata (including 3-D neuronal coordinates, stimulus information, eye tracking data, and running speed) and packaged into NWB files. Each such file is intended to be a complete, self-contained data record for one experimental session in one animal. NWB files are uploaded to three different locations in the cloud: The AWS Registry of Open Data, the DANDI Archive, and the Allen Institute data warehouse (accessed via the AllenSDK, a Python API for searching for relevant sessions and downloading data). Raw data is also uploaded to the AWS Registry of Open Data. End users can either analyze data in the cloud (for public cloud datasets) or download data for local analysis.

## Three Families of Use Cases

The first round of 2-photon calcium imaging data was released in July 2016, followed by 3 subsequent releases that expanded the dataset (green triangles in **Figure 3**). The Neuropixels dataset became available in October 2019 (yellow triangle in **Figure 3**). There are now 104 publications or preprints that reuse these two datasets, with first authors at fifty unique institutions (see **Supplemental File**). This demonstrates the broad appeal of applying a survey-style approach to the domain of *in vivo* neurophysiology

We found three general use cases of Allen Brain Observatory data in the research community (**Figure 4**):

(i) Generating novel discoveries about brain function
(ii) Validating new computational models and algorithms
(iii) Comparing with experiments performed outside the Allen Institute

Below, we highlight some examples of these three use cases, for both the 2-photon calcium imaging and Neuropixels datasets. All these studies were carried out by groups external to the Allen Institute, and frequently without any interaction from us, speaking to the ease with which data can be downloaded and analyzed.

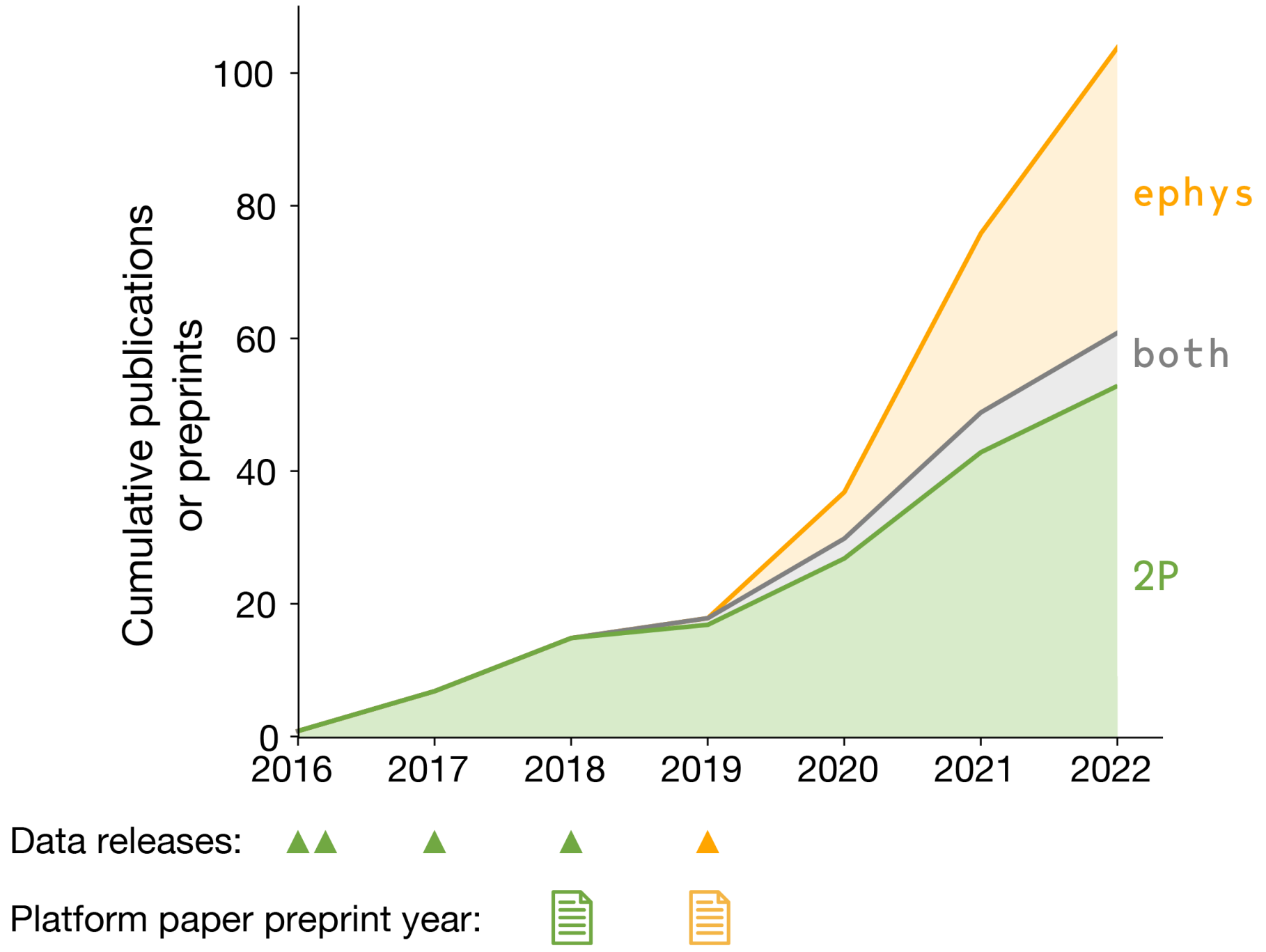

**Figure 3 – Data reuse over time.** Cumulative number of papers or preprints that include novel analysis of Allen Brain Observatory Visual Coding surveys. Triangles indicate the years in which new data was made publicly available. Paper icons indicate the years in which the Allen Institute preprints describing the dataset contents and initial scientific findings were posted.

**Generating novel discoveries**

Sweeney and Clopath (2020) used Allen Brain Observatory 2-photon imaging data to explore the stability of neural responses over time. They had previously found that neurons in a recurrent network model with high inherent plasticity had more variability in their stimulus selectivity than those with low plasticity. They had also found that neurons with high inherent plasticity had higher population coupling. To examine whether these were related, they here used analyzed real calcium-dependent fluorescence traces to examine whether population coupling and response variability were tethered. The authors found that, indeed, population coupling is correlated with the change in orientation and direction tuning of neurons over the course of a single experiment, an unexpected result linking population activity with individual neural responses.

Bakhtiari et al. (2021) examined whether a deep artificial neural network (ANN) could model both the ventral and dorsal pathways of the visual system in a single network. They trained two networks, one with a single pathway and the other with two parallel pathways, using a Contrastive Predictive Coding loss function. Comparing the representations of these networks with the neural responses in the 2-photon imaging dataset, they found that the single pathway produced ventral-like representations but failed to capture the activities of the dorsal areas. The parallel pathway network, though, induced distinct representations that mapped onto the ventral/dorsal division. This work is an illustration of how large-scale data can guide the development of modeling and artificial intelligence, and, conversely, how those approaches can inform our understanding of cortical function.

Fritsche et al. (2022) analyzed the time course of stimulus-specific adaptation in 2,365 neurons in the Neuropixels dataset and discovered that a single presentation of a drifting or static grating in a specific orientation leads to a reduction in the response to the same visual stimulus up to eight trials (22 seconds) in the future. This stimulus-specific, long-term adaptation persists despite intervening stimuli, and is seen in all six visual cortical areas, but not in visual thalamic areas (LGN and LP), which returned to baseline after one or two trials. This is a remarkable example of a discovery that was not envisioned when designing our survey, but for which our stimulus set was well suited.

Nitzan et al. (2022) took advantage of the fact that every Neuropixels insertion targeting visual cortex and thalamus also passed through the intervening hippocampus and subiculum. They analyzed the local field potential from these electrodes to detect the onset of sharp-wave ripples, fast oscillations believed to mediate offline information transfer out of the hippocampus (Girardeau and Zugaro, 2011). They found that sharp-wave ripples coincided with a transient, cortex-wide increase in functional connectivity with the hippocampus. Although the Allen Brain Observatory experiments were not originally designed to test hypotheses of hippocampal function, the Neuropixels dataset turned out to be attractive for understanding the interactions between this structure and visual cortical and thalamic regions.

**Validating models and algorithms**

Many researchers have used the numerous and diverse fluorescence movies in the 2-photon imaging dataset to validate image processing algorithms. As the different transgenic lines used in the dataset target different populations of neurons, they have different labeling densities. As a result, there are some very sparse movies with only a dozen neurons within the field of view, and others with up to ~400 neurons. This makes the dataset a rich resource for benchmarking methods for cell segmentation (Bao et al., 2021; Inan et al., 2021; Kirschbaum et al., 2020; Petersen et al., 2018; Soltanian-Zadeh et al., 2019), matching neurons across multiple sessions (Sheintuch et al., 2017), and removing false transients in the fluorescence traces (Bao et al., 2022; Gauthier et al., 2022).

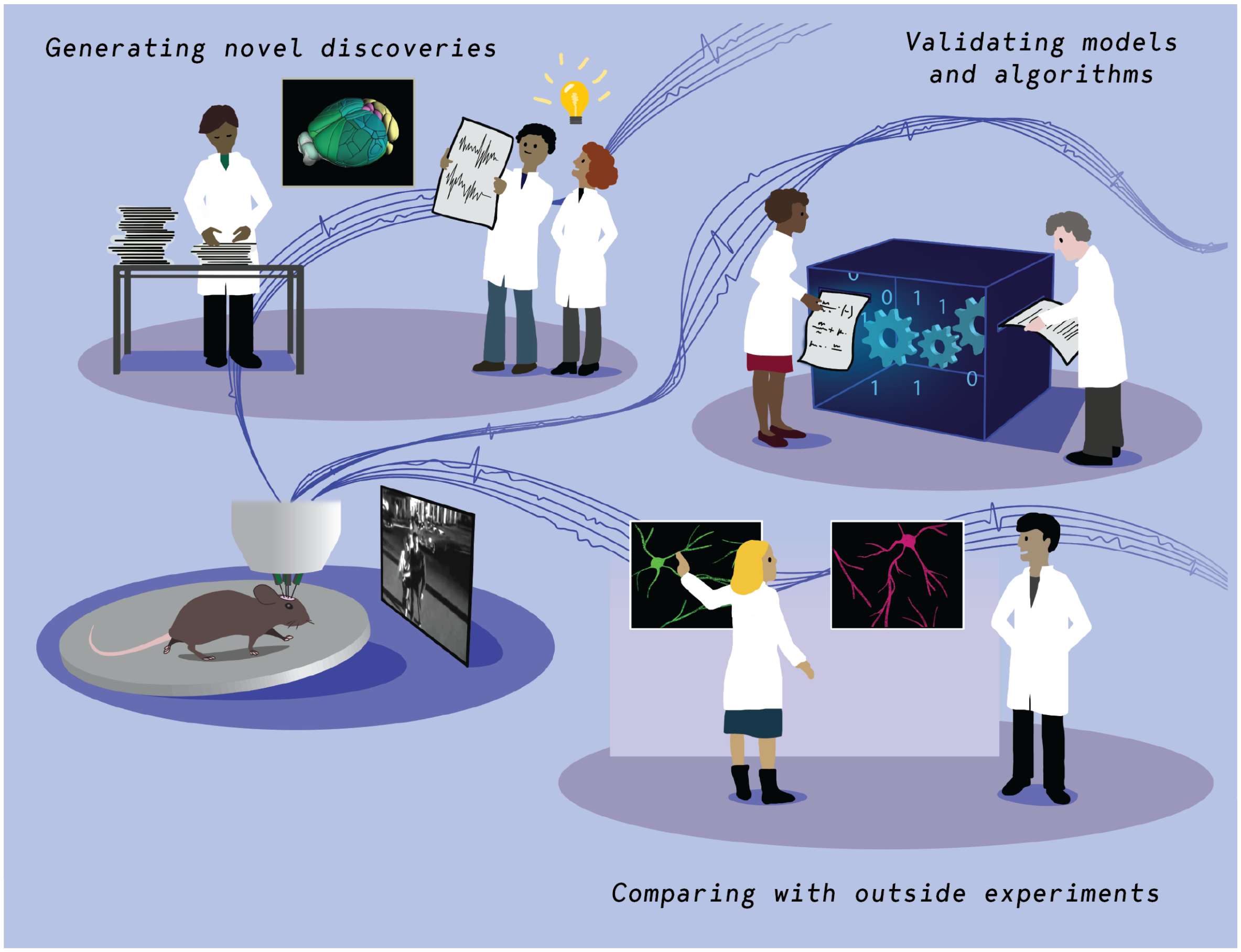

**Figure 4 – Primary use cases for Allen Brain Observatory data.** *Bottom left*: The Allen Institute collects cellular-level physiology data from awake mice passively exposed to a battery of visual stimuli. *Upper left*: Researchers mine the data to produce novel discoveries. *Upper right*: Researchers use the data to validate and characterize computational algorithms. *Lower right*: Researchers compare findings from datasets produced in-house to those from the publicly available data.

Montijn et al. (2021) used the Neuropixels survey to showcase a novel method for identifying statistically significant changes in neural activity. Their method, called *ZETA* (Zenith of Event-based Time-locked Anomalies), detects whether a cell is responsive to stimulation without the need to tune parameters, such as spike bin size. As an example, they analyze the "optotagging" portion of the Neuropixels experiments carried out in Vip-Cre x ChR2 mice, involving the activation of Vip+ interneurons with brief pulses of blue light. Intended to aid in the identification of genetically defined cell types at the end of each recording session, the authors show how these recordings can be exploited to test the network-level impact of triggering a particular class of interneurons. ZETA identifies not only Vip+ neurons that are directly activated by the light pulses, but also nearby cortical neurons that are inhibited on short timescales and disinhibited over longer timescales.

Buccino et al. (2020) used raw data from the Neuropixels survey to validate *SpikeInterface,* a Python package that runs multiple spike sorting algorithms in parallel and compares their outputs. We originally performed spike sorting with one such algorithm, *Kilosort 2* (Pachitariu et al., 2016). The authors of this paper used SpikeInterface to compare the performance of Kilosort 2 and five additional algorithms. In one example session, over 1000 distinct units were detected by only one sorter, while only 73 units were detected by five or more sorters. At first glance, this finding seems to indicate a high level of disagreement among the algorithms. However, when comparing these results with those from simulations, it became clear that the low-agreement units were mainly false positives, while the true positive units were highly consistent across algorithms. This finding, and the SpikeInterface package in general, will be essential for improving the accuracy of spike sorting in the future.

**Comparing with other datasets**

Kumar et al. (2021) used supervised and semi-supervised learning algorithms to classify cortical visual areas based on either spontaneous activity or visually evoked responses. Cortical visual areas, defined based on retinotopic maps, are thought to serve distinct visual processing functions. Rather than compare tuning properties of neurons across the areas, as many studies (including our own) have done, the authors trained classifiers to successfully determine the area membership and boundaries from the neural responses to visual stimuli. They compared the performance of these algorithms for their own wide-field imaging dataset with our 2-photon imaging dataset. This provides an extension and validation of their results to conditions in which single-cell responses are available.

Muzzu and Saleem (2021) performed electrophysiological recordings in the mouse cortex to examine "mismatch" responses, where neurons respond to differences in visual cue and motor signals from running. The authors argued that these responses derive from visual features rather than the mismatch, showing that these perturbation responses might be explained by preferential tuning to low temporal frequencies. The authors use our 2-photon imaging dataset to demonstrate a difference in temporal frequency tuning across cortical layers, with neurons in superficial layers being tuned to lower frequencies, supporting the fact that mismatch responses are predominantly observed in superficial layers. While this use case is perhaps one of the simplest, it is an elegant demonstration of gaining validation for implications that emerge in one's own experiments.

Stringer et al. (2021) compared spiking activity from the Neuropixels dataset to calcium-dependent fluorescence changes recorded in their laboratory. Their analysis focused on the precision with which the orientation of static gratings can be decoded from activity in visual cortex. Using their own 2-photon calcium imaging dataset that consisted of up to 50,000 simultaneously recorded neurons, they found that it was possible to use neural activity to discriminate orientations that differ by less than 0.4°, about a factor of 100 better than reported behavioral thresholds in mice. As an important control, they showed that the trial-to-trial variability in evoked responses to static gratings was nearly identical between their 2-photon data and our Neuropixels electrophysiology data, indicating that their main result was not likely to depend on the recording modality. This use case is noteworthy because the preprint containing this comparison appeared less than a month after our dataset became publicly available.

Schneider et al. (2021) directly compared the Allen Neuropixels dataset with Neuropixels recordings from LGN and V1 carried out locally. They first analyzed gamma-band coherence between these two structures in the Allen Brain Observatory dataset and found evidence in support of their hypothesis that inter-regional coherence is primarily driven by afferent inputs. This contrasts with the "communication through coherence" hypothesis (Fries, 2015), which posits that pre-existing inter-regional coherence is necessary for information transfer. They then performed a separate set of Neuropixels recordings in which they found that silencing cortex (via optogenetic activation of somatostatin-positive interneurons) did not change the degree of coherence between LGN and V1, indicating that V1 phase-locking is inherited from LGN, further supporting their hypothesis. This is an insightful example of how a survey dataset can be used to test a hypothesis, followed by a set of more specific follow-up experiments that refine the initial findings.

### Use in education

Data from these surveys has also been used in a variety of educational contexts. Many computational neuroscience summer courses have presented these data as potential source of student projects. This includes the Allen Institute's own *Summer Workshop on the Dynamic Brain* as well as the Cold Spring Harbor *Neural Data Science* and *Computational Neuroscience: Vision* courses; Brains, Minds, and Machines Summer Course at the Marine Biological Laboratory; and the Human Brain Project Education Programme. Indeed, in some cases these projects have led to publications (Christensen and Pillow, 2022; Conwell et al., 2022). Beyond these summer courses, these datasets are discussed in undergraduate classrooms, enabling students to learn computational methods with real data rather than toy models. This includes classes at the University of Washington, Brown University, and the University of California, San Diego.

## Discussion

Although it is too early to assess the long-term relevance of the first two Allen Brain Observatory datasets, the more than 100 publications that mined this data over the last six years testify to its immediate impact. Our data has been used for a wide array of applications, many of which we did not envision when we designed the surveys. We attribute this success to several factors, including the scale of the dataset (tens of thousands of neurons across hundreds of subjects), our extensive curation and documentation efforts (in publications, white papers, and websites), a robust software kit for accessing and analyzing the data (the AllenSDK), and a well-organized outreach program (involving tutorials at conferences and a dedicated summer workshop).

One key lesson we learned is to facilitate different types of data reuse, as illustrated by the examples above. While many users primarily care about spike times or fluorescence traces, others require raw data. Because of this, it was imperative for us to provide access to both (**Figure 2**). Sharing the data in a way that is flexible and does not constrain which questions can be addressed is paramount for facilitating reuse.

We hope to see the sharing of both raw and processed cellular physiology data soon become ubiquitous. However, we know that our surveys were contingent on the efforts of a large team, including scientists from multiple disciplines, hardware and software engineers, research associates, and project managers. Assembling similar resources is untenable for most academic labs. Fortunately, there are ongoing developments that will lower the barriers to sharing and re-using data: *increased standardization* and *cloud-based analysis tools*.

**Increased standardization**

The success of data reuse rests on the FAIR Principles: data must be Findable, Accessible, Interoperable, and Reusable (Wilkinson et al., 2016). In other words, prospective analysts must be able to easily identify datasets appropriate to their needs and know how to access and use the data assets. This is best accomplished if data is stored in standardized formats, with common conventions for rich metadata and easy-to-use tools for search and visualization.

The Allen Institute has invested heavily in developing and promoting *Neurodata Without Borders* (NWB) as a standard data model and interchange format for neurophysiology data (Rübel et al., 2022; Teeters et al., 2015). NWB has been criticized for being both too restrictive (as it often takes a dedicated programmer to generate format-compliant files from lab-specific data) and not restrictive enough (as it does not enforce sufficient metadata conventions, especially related to behavioral tasks). Nevertheless, there are overwhelming advantages to having common, language-agnostic formatting conventions across the field. Building a rich ecosystem of analysis and visualization tools based on NWB will incentivize additional labs to store their data in this format and even to directly acquire data in NWB files to make data immediately shareable (this is already possible for electrophysiological recordings using the Open Ephys GUI; Siegle et al., 2017).

Standardized metadata conventions are also critical for promoting data reuse. Our surveys are accompanied by extensive white papers, code repositories, and tutorials that detail the minutiae of our methods and tools, beyond the standard "Methods" section in publications (see **Appendix** for links). For the community at large, a more scalable solution is needed. Standardized and machine-readable metadata needs to extend beyond administrative metadata (describing authors, institutions and licenses) to include thorough and detailed experimental conditions and parameters in a self-contained manner. As data sharing becomes more widespread, standardization of metadata will be particularly important for enhancing "long tail" effects. Most publicly available data will come from more focused studies that may not adhere to the same quality control standards of large-scale surveys. To avoid a situation in which smaller-scale datasets are overlooked, all these studies should be indexed by a single database that can be filtered by relevance, making it much easier for researchers to decide what data is appropriate for their needs.

The recently launched DANDI Archive attempts to address this issue by enforcing the use of NWB for all shared datasets (Rübel et al., 2022). In the two years since the first dataset was uploaded, the archive now hosts more than 100 NWB-formatted datasets accessible via download links, a command-line interface, or within a cloud-based JupyterHub environment. While the adoption of consistent data formatting conventions is a welcome development, there are also benefits to greater standardization of protocols, hardware, and software used for data collection. One way this can be achieved is through coordinated cross-laboratories experiments, such as those implemented by the International Brain Laboratory (IBL), a consortium that is surveying responses across the entire mouse brain in a visual decision-making task (Abbott et al., 2017; Ashwood et al., 2022).

To encourage data sharing, the field also needs greater standardization in the way data mining is tracked and credited. Digital object identifiers (DOIs) are an essential first step; we regret not making them an integral part of the Visual Coding data releases. However, they have not solved the problem of discovering reuse, as they are not always included in publications. It is more common to include a reference to the original paper in which the dataset was described, but this makes it difficult to distinguish instances of reuse from other types of citations. Currently the onus is on those releasing the data to keep track of who accesses it. There is an acute need for a solution, especially if funding agencies and hiring and promotion committees want to reward data sharing efforts (which we strongly believe they should). To take one example, the **cai-1** GCaMP calibration dataset from the Svoboda Lab at HHMI Janelia Research Campus (GENIE Project, 2015) only has 5 citations tracked in Google Scholar. Yet a deeper dive into the literature reveals that this dataset has been reused in a wide range of publications and conference papers that benchmark methods for inferring spike rate from calcium fluorescence signals, of which there are likely over 100 in total. Many of these papers only cite the original publication associated with this dataset (Chen et al., 2013), refer to the repository from which the data was downloaded (CRCNS), or do not cite the data source at all. The lack of an agreed-upon method for citing datasets (like we have for journal articles) is a huge loss for the community, as it hinders our ability to give appropriate credit to those responsible for collecting widely used datasets.

**Cloud-based analysis tools**

To enable more efficient data mining, end users should ideally not need to download data at all. This is particularly true as the volume of data keeps on growing (for example, a single Allen Brain Observatory Neuropixels session generates about 1.2 TB of raw data). Therefore, the goal should be to bring users to the data, rather than the data to users.

Generic analysis tools, such as Amazon's SageMaker and Google's Colab, already make it possible to set up a familiar coding environment in the cloud. However, we are most excited about tools that lower the barriers and the costs of cloud analysis. Some of the most promising tools include DataJoint (Yatsenko et al., 2015), DandiHub, NeuroCAAS (Abe et al., 2022), Binder, and CodeOcean (many of which are built on top of the powerful Jupyter platform). All of these are aimed at improving the reproducibility of scientific analyses, while shielding users from the details of configuring cloud services.

Cloud-based analysis is not a panacea. Although individual tools can be vendor-agnostic, there will be a push to centralize around a single cloud platform, given the high cost of data egress. This could lead to a single company monopolizing the storage of neurophysiology data; it would therefore be prudent to invest in a parallel distribution system that is controlled by scientists (Saunders, 2022). In addition, it is (perhaps not surprisingly) notoriously easy to provision expensive cloud computing resources; a single long-running analysis could exhaust a lab's entire annual budget if they are not careful. Many labs will not be willing to take this risk, meaning that additional services will be needed to prevent cost overruns, or to provide insurance when they do occur. Despite these drawbacks, we believe that a move to cloud-based analysis will be essential for reducing the friction involved in adopting new datasets. We plan to move toward supporting a cloud-native sharing model more directly in our upcoming data releases.

**Fostering a culture of data reuse**

The value of open data is best realized when it is conveniently accessible. Whether this involves new discoveries or comparing results across studies, data mining is vital for progress in neuroscience, especially as the field as a whole shifts toward more centralized "Observatories" for mice and non-human primates (Koch et al., 2022). The BRAIN Initiative has invested considerable resources in advancing instruments and methods for recording large number of neurons in more sophisticated behavioral contexts. Yet the analytical methods for understanding and interpreting large datasets are lagging, as many of our theoretical paradigms emerged from an era of small-scale recordings (Urai et al., 2022). In order to develop theories that can explain whole-brain cellular data, it is critical to maximize data reuse.

This poses a set of challenges. First and foremost, as mentioned above, we need increased compliance with standards and enhanced tooling around those standards. Scientists cannot afford to spend time trying to understand obscure data formats or deciphering which experimental conditions were used; analyzing data collected by another lab needs to be straightforward and intuitive. We must be able to focus on actual biology rather than on the technical challenges of sharing and accessing data. Secondly, we need additional guardrails that can identify and prevent "inappropriate" reuse, such as analyses and interpretations that fail to account for the limitations of an experimental paradigm. For instance, we have shown that a naïve comparison of cellular properties measured in the same visual areas across our Neuropixels and 2-photon calcium imaging datasets reveals substantial discrepancies (Siegle et al., 2021b). These can only be reconciled by accounting for the bias inherent in each recording modality, as well as the data processing steps leading to the calculation of functional metrics. Effective data reuse requires that we, as a field, focus more of our energies on better communicating these important technical factors.

Neuroscientists have traditionally been taught to address questions by collecting new data. As data sharing becomes more prevalent, neuroscientists' first instinct should instead be to search for existing data that may offer insights into the problem at hand, whether or not it was originally intended for this purpose. Even in situations where the "perfect" dataset does not yet exist, it is likely that researchers can exploit available data to refine a broad question into one that is more focused, and thus experimentally more tractable. Just as young scientists are trained to discover, interpret, and cite relevant publications, it is imperative that they are also taught to effectively identify, evaluate, and mine open datasets.

## Acknowledgements

We thank all the members of Transgenic Colony Management, Lab Animal Services, Neurosurgery & Behavior, Imaging and Neuropixels Operations Teams, Materials & Process Engineering, Information Technology, and Program Management that cared for and trained the animals, built and staffed the instruments, processed the brains, and wrangled the data streams. We thank Allan Jones for providing an environment that nurtured our efforts and the Allen Institute founder, Paul G. Allen, for his vision, encouragement, and support. This research was funded by the Allen Institute. We thank Amazon Web Services for providing free cloud data storage as part of the Open Data Registry program. We thank Bénédicte Rossi for creating the illustration in Figure 4. We thank Karel Svoboda, David Feng, Jerome Lecoq, Shawn Olsen, Stefan Mihalas, Anton Arkhipov, and Michael Buice for feedback on the manuscript.

## Appendix – Web Resources for Allen Brain Observatory Visual Coding Datasets

*White papers describing the surveys*
2P – http://help.brain-map.org/display/observatory/Documentation
Neuropixels – https://portal.brain-map.org/explore/circuits/visual-coding-neuropixels

*Code repositories*
AllenSDK – https://github.com/alleninstitute/allensdk
2P – https://github.com/AllenInstitute/visual_coding_2p_analysis
Neuropixels – https://github.com/AllenInstitute/neuropixels_platform_paper

*Tutorials*
2P – https://allensdk.readthedocs.io/en/latest/brain_observatory.html
Neuropixels – https://allensdk.readthedocs.io/en/latest/visual_coding_neuropixels.html